# Unveiling the critical role of interfacial strain in adjusting electronic phase transitions in correlated vanadium dioxide

*Xuanchi Zhou* [1, 2] *, *Xiaohui Yao* [1], *Xiaomei Qiao* [1]

[1] *Key Laboratory of Magnetic Molecules and Magnetic Information Materials of Ministry of Education & School of Materials Science and Engineering, Shanxi Normal University, Taiyuan, 030031, China*

[2] *Research Institute of Materials Science, Shanxi Key Laboratory of Advanced Magnetic Materials and Devices, Shanxi Normal University, Taiyuan 030031, China*

*Authors to whom correspondence should be addressed: *xuanchizhou@sxnu.edu.cn* (X. Zhou).

**Abstract**

Thermally activated abrupt switching between localized and itinerant electronic states during the insulator-metal transition (IMT) in correlated oxide systems serves as a powerful platform for exploring exotic physical phenomena and device functionality. One ongoing focal challenge lies in the realization of the broadly tunable IMT property in correlated system, to satisfy the demands of practical applications across diverse environments. Here, we unveil the overwhelming advantage associated with interfacial strain in bridging the bandwidth and band-filling control over the IMT property of $VO_2$. Tailoring the orbital overlapping through strain-mediated bandwidth control enables a widely tunable thermally-driven IMT property in $VO_2$. Benefiting from adjustable defect dynamics, filling-controlled Mott phase modulations from electron-localized $t_{2g}^1 e_g^0$ state to electron-itinerant $t_{2g}^{1+\Delta} e_g^0$ state through oxygen vacancies can be facilitated by using *in-plane* tensile distortion, overcoming the high-speed bottlenecks in iontronic devices. Defect-engineered electronic phase transitions are primarily governed by the electron filling in $t_{2g}$ band of $VO_2$, showcasing a definitive relationship with the incorporated defect concentration. Our findings provide fundamentally new insights into the on-demand design of emergent electronic states and transformative functionalities in correlated oxide system by unifying two fundamental control paradigms of bandwidth and band-filling control.

1. **Introduction**

The intricate interplay among charge, orbital, lattice, and spin degrees of freedom in correlated oxide system gives rise to exotic physical phenomena, transcending traditional semiconductors.[1-3] As a representative case, insulator-metal transition (IMT), driven by electron-electron correlations and/or electron-lattice interactions, underpins rich correlated physics and multidisciplinary device applications by leveraging an abrupt switch from localized to itinerant electronic behavior.[4-6] The precise control over the IMT behavior of correlated oxides enables the possibility in exploring exotic electronic states and device functionalities, a focal point in correlated electronics. In recent years, the incorporation of mobile ions (e.g., protons or oxygen vacancies) has inaugurated a new paradigm to regulate the IMT functionality of correlated oxides in a more reversible and controllable pathway.[7-10] Specifically, oxygen defects can significantly modify the electronic band structure via introducing the electron carriers into the conduction band, while driving novel structural transformations through oxygen vacancy ordering.[11, 12] The unconventional superconductivity in infinite-layer nickelates through topochemically reducing hole-doped perovskite parents exemplifies the irreplaceable role of oxygen vacancy ordering in discovering new quantum states.[2] Analogously, hydrogen doping can reversibly trigger magnetoelectric phase modulations in correlated oxides, from which novel physical phenomena such as electronic antidoping have emerged.[13-15] This breakthrough not only extensively extends magnetoelectric phase diagram, with new electron phases and magnetic ground states, but also boosts device applications in electrochemical transistors,[16] energy conversions [17, 18] and artificial intelligence.[19, 20]

Beyond filling-controlled Mott physics, interfacial strain serves as an alternative tuning knob for modulating the IMT functionality in correlated oxides through bandwidth control. Imparting the interfacial strain to correlated oxides enables a widely tunable IMT behavior through flexibly adjusting the *d*-orbital bandwidth and orbital hybridization.[21] For example, in metastable *Re*NiO$_3$ system with a distorted perovskite structure, *in-plane* tensile distortion reduces the Ni-O-Ni bond angle from 180 º to depress the orbital hybridization between Ni-3*d* and O-2*p* orbitals, elevating the resultant transition temperature ($T_{IMT}$).[22] Benefiting from the robust ability to manipulate the migration kinetics of mobile ions,[23] interfacial strain provides a powerful pathway for designing resultant physicochemical functionality in correlated oxides through ionic evolution, from the microscopic perspective (Figure 1a). By introducing an additional degree of freedom, interfacial strain creates a powerful platform for functionalizing correlated oxides, enabling the rational design of novel magnetoelectric properties [24] and advanced energy materials like electrocatalysts [25, 26] and solid oxide fuel cells [27]. Consequently, strain engineering endows with the unique capability to not only design the thermally-driven IMT via bandwidth control but also to tailor band-filling-controlled phase transformations by modifying the ionic mobility. Such the overwhelming advantage associated with strain engineering over traditional tuning strategies extends the horizons in on-demand materials design for unlocking unprecedented functionalities, surpassing the high-speed challenge in iontronic devices.

Here, correlated $VO_2$ is delicately selected as a model system to demonstrate the robust capability of interfacial strain in adjusting thermally-driven and defect-mediated electronic phase transitions. As a typical $3d^1$-orbital correlated oxide, $VO_2$ undergoes a pronounced IMT functionality at the $T_{IMT}$ of 341 K that is readily tunable by using the interfacial strain. Benefiting from the lattice mismatch between rutile $VO_2$ and $TiO_2$, the compressive distortion along the *c*-axis direction of rutile $VO_2$, parallel to the V-V pairs, results in an enhanced orbital overlapping between V-$3d$ and O-$2p$ orbitals and a wider bandwidth of V-$3d$ orbital, thereby reducing the $T_{IMT}$. Beyond that, *in-plane* tensile distortion facilitates defect-mediated filling-controlled phase transformations via adjusting oxygen defect dynamics, during which the introduction of electron carriers tends to occupy the low-energy $t_{2g}$ band. Our work establishes the interfacial strain as a versatile gateway to manipulate the IMT functionality of correlated $VO_2$ system, triggered by either critical temperature or defect engineering, addressing the critical bottleneck in developing high-speed iontronic devices.

## 2. Experimental Section

The $VO_2$ thin films were epitaxially grown on single-crystalline $TiO_2$ (001) substrates via using laser molecular beam epitaxy (LMBE). Prior to the film deposition, the growth chamber was evacuated to a base pressure lower than $10^{-5}$ Pa. The $VO_2$ films were deposited at an optimized substrate temperature of 300 °C under an oxygen partial pressure of 1.5 Pa, with a target-substrate distance of 45 mm and a laser fluence of 1.0 J·cm$^{-2}$. Following the film deposition, the $VO_2/TiO_2$ (001) heterostructures were cooled down to the room temperature under the same oxygen pressure. To controllably introduce the oxygen deficiency, the as-grown $VO_2/TiO_2$ (001) samples were subjected to high-vacuum annealing (e.g., $P_{O_2} \sim 1 \times 10^{-5}$ Pa) at 300 ºC for different annealing duration.

The crystal structure of the $VO_2/TiO_2$ (001) heterostructures was characterized by using X-ray diffraction (XRD, Rigaku Ultima IV) and high-resolution transmission electron microscopy (HRTEM, FEI Tecnai G2 S-TWIN). Epitaxial growth of the $VO_2$ film was verified by using the reciprocal space mapping (RSM, RIGAKU SmartLab). Chemical environments of oxygen-deficient $VO_{2-x}$ were analyzed by the X-ray photoelectron spectroscopy (XPS, Thermo K-Alpha). Soft X-ray absorption spectroscopy (sXAS) measurements were performed at beamline BL08U1A of the Shanghai Synchrotron Radiation Facility (SSRF) to probe the electronic structure of oxygen-deficient $VO_{2-x}$. Electrical transport properties were measured using a physical property measurement system (PPMS, Quantum Design), while the room-temperature material resistance was evaluated with a commercial Keithley 4200 setup.

## 3. Results and discussion

Strain-mediated bandwidth control and defect-engineered band-filling control represent two distinct pathways for precisely modulating the IMT property in correlated $VO_2$ system, from the perspective of Mott physics. By leveraging the capability to control oxygen defect dynamics, interfacial strain is expected to provide a unified framework that not only bridges these central Mott paradigms but also surmounts the primary bottleneck to realizing high-speed iontronics. To address the above central concept, rutile $VO_2$ films were epitaxially deposited on rutile $TiO_2$ (001) substrate using the laser molecular beam epitaxy (LMBE) technique, leveraging their *in-plane* lattice mismatching to impose controlled strain states. To identify the heteroepitaxial growth of as-grown $VO_2$ (15 nm)/$TiO_2$ (001) heterostructures, high-resolution transmission electron microscopy (HRTEM) equipped with energy dispersion spectrum (EDS) was performed, as the results shown in Figures 1b-1d. Specifically, the low-magnification cross-sectional HRTEM image demonstrates the continuous and uniform growth of $VO_2$ film on $TiO_2$ (001) substrate (Figure 1b). Furthermore, the rutile-on-rutile epitaxial growth of $VO_2$ films deposited on *c*-plane $TiO_2$ substrate is directly verified by the atomic coherence in the heterointerface region for as-grown $VO_2$/$TiO_2$ (001) heterostructures (Figures 1c and S1). The EDS elemental mappings reveal a chemically distinct boundary, albeit with a ~2-3 nm graded interfacial region resulting from the interdiffusion of $Ti^{4+}$ ions (Figure 1d).

Further consistency in the epitaxial relationship of $VO_2$/$TiO_2$ (001) heterostructure is demonstrated by reciprocal space mapping (RSM) in Figure 2a, where the identical *in-plane* vector (e.g., $Q_\parallel$) is identified for the grown $VO_2$ film and $TiO_2$ substrate. In addition, the *cross-plane* vector (e.g., $Q_\perp$) for $VO_2$ film is enlarged with respect to the counterpart of $TiO_2$ substrate, confirming an *in-plane* biaxial tensile distortion. This observation is consistent with the smaller *a*-axis lattice constant ($a_0$) of the $VO_2$ film ($a_{0, film}$ = 4.5546 Å) compared to the $TiO_2$ substrate ($a_{0, sub.}$ = 4.5937 Å), as expected for a coherently strained heterointerface, which induces an *in-plane* tensile strain on the grown $VO_2$ film. Nevertheless, further elevating the film thickness of $VO_2$ to 50 nm instead leads to the partial strain relaxation of the grown $VO_2$/$TiO_2$ heterostructures due to formation of dislocations in the heterointerfacial region, as indicated by the different $Q_\parallel$ between $VO_2$ film and $TiO_2$ substrate (Figure 2b). As expected, the peak intensity in the RSM mapping for the 15 nm-thick $VO_2$ film is significantly lower than that of the 50 nm counterpart, attributed to the weaker scattering signal from the thinner layer. Therefore, the interfacial strain states in the grown $VO_2$ films can be effectively engineered by simply altering the film thickness, in accordance with previous reports. [21, 28] In addition, the film thickness for the grown $VO_2$ films is precisely identified by using atomic force microscope (AFM) (Figure 2c), while respective AFM topography reveals a relative smoothy film surface with the root-mean-square roughness ($R_q$) values around 0.2 nm (Figure S2).

On the basis of the delicate design in the *in-plane* tensile strain states via altering the film thickness, the thermally activated IMT behavior of $VO_2$ is poised to be effectively adjusted via bandwidth control. This understanding is clearly demonstrated by comparing the temperature dependence of material resistivity ($\rho$-$T$ tendency) for the $VO_2$ films deposited on $c$-plane $TiO_2$ and $Al_2O_3$ substrates (Figure 2d). It is well-known that the V-V chains, aligned parallel to the $c$-axis of rutile $VO_2$, govern the V-3$d$ orbital overlapping, offering the fertile ground for tailoring the IMT behavior through bandwidth control. By strong contrast, the domain-matching epitaxy of rutile $VO_2$ film on $c$-plane hexagonal $Al_2O_3$ substrate yields a strain-relaxed state,[29] aligning well with the standard $T_{IMT}$ to the bulk value. It is worthy to point out a counterintuitive negative value of temperature coefficient of resistance (*TCR*) in metallic phase of the grown $VO_2$ film on $c$-plane $Al_2O_3$ template, deviating from conventional metallic transport behavior, which also starkly differs from the case of $VO_2/TiO_2$ heterostructure. This abnormal phenomenon is associated with the possibly existing twin variants in the $VO_2$ film arising from the domain-matching epitaxy, leading to the residual insulating phase that cannot completely transits into the metallic state above the $T_{IMT}$. Imparting the tensile distortion to $VO_2$ along the *in-plane* direction extensively reduces the resultant $T_{IMT}$ to 291.59 K. This observation is in agreement with the previous reports associated with *in-plane* tensile strained $VO_2$, in which case the compressive distortion along the $c$-axis direction of rutile $VO_2$ via the Poisson effect gives rise to an intensified V-3$d$ and O-2$p$ orbital hybridization and a wider bandwidth.[21] Similarly, the $T_{IMT}$ realized in 50 nm-thick $VO_2$ film deposited on the $c$-plane $TiO_2$ substrate gradually recovers toward the bulk value, consistent with the partial strain relaxation. The emergence of step-like IMT behavior for 50 nm-thick $VO_2$, together with the depression in transition abruption, may be related to the strain relaxation [30] or separate phase transition, worthy of future exploration. Consequently, the interfacial strain emerges as an effective tuning knob for modifying the IMT property of correlated $VO_2$ via bandwidth control, rendering a tunable IMT behavior.

Apart from strain-mediated bandwidth control, the introduction of each oxygen vacancy readily donates two electron carriers into the unoccupied conduction band that modifies the electronic band structure via band-filling control, as described by $O_{O_x} \rightarrow V_O^{\bullet\bullet} + 2e^- + \frac{1}{2}O_2$ (g) (Figure 3a). To controllably introduce oxygen vacancies, the grown $VO_2$ films were annealed at 300 °C for 1 h under a high-vacuum atmosphere with a $P_{O_2}$ below $1 \times 10^{-5}$ Pa, which facilitates the oxygen desorption away from the $VO_2$ lattice. The diffraction peak representing the (002) plane for the grown $VO_2$ film appears in close proximity to the one for (002) plane of $TiO_2$ substrate, unveiling an *out-of-plane* preferential film growth (Figure 3b). The introduction of oxygen deficiency results in the lattice expansion of $VO_2$ film along the *out-of-plane* direction, evidenced by the leftwards shift of characteristic diffraction peak towards a lower angle. However, the *in-plane* lattice of $VO_2$ film is expected to be tightly locked by epitaxial $TiO_2$ template, with the anisotropy in defect-driven structural evolution.[23]

Notably, the lattice expansion for VO$_2$ film triggered by the oxygen deficiency can be intensified by imparting the *in-plane* tensile distortion, in which the lattice expansion in 15 nm tensile-distorted VO$_2$ extensively exceeds that of 50 nm strain-relaxed counterpart by one order of magnitude.

Oxygen vacancies introduce additional electron carriers to occupy the low-energy empty states in the conduction band of VO$_2$ that enlarges the relative stability in the metallic phase relative to the insulating phase. This understanding is clearly demonstrated by the *ρ-T* tendencies in Figure 3c, wherein the $T_{IMT}$ for 50 nm oxygen-deficient VO$_{2-x}$ film with partial strain relaxation is reduced from 291.59 K to 250.38 K, accompanied by a depressed transition sharpness. The magnitude of $T_{IMT}$ was operatively defined as the average of the critical points in the heating and cooling *TCR-T* curves, with each critical temperature identified from the corresponding local maximum (Figure S3). Analogously, defect-mediated electron doping substantially reduces the $T_{IMT}$ for strain-relaxed VO$_2$ film deposited on *c*-cut Al$_2$O$_3$ substrate from 349.95 K to 299.79 K (Figure S4). The above findings are consistent with the metallization of VO$_2$ through defect engineering that enlarges the relative stability in metallic phase to reduce the resultant $T_{IMT}$.[12]

Nevertheless, oxygen deficiency can trigger the carrier delocalization of 15 nm tensile-distorted VO$_2$ under identical annealing conditions, with the complete depression in the IMT property being observed. This observation underscores the critical role of interfacial strain in accelerating both the electronic phase transition and structural evolution. The tunability in defect-driven electronic phase transition under *in-plane* tensile distortion is further confirmed by altering the film thickness and annealing period (Figures S5-S6). Remarkably, defect-associated electronic phase modulations showcase a reversible trend towards the pristine state under oxidative annealing at 200 ºC for 2 h, despite the retained oxygen vacancies in the deeper layer hinder the complete restoration. Such the recovery process driven by oxidative annealing is analogously expediated by imparting *in-plane* tensile distortion, which effectively restores the IMT characteristics in tensile-distorted VO$_2$, contrasting with similar transport behavior in strain-relaxed VO$_2$. In previous work, *in-plane* tensile distortion was demonstrated to manipulate microscopic hydrogen distribution in metastable perovskite NdNiO$_3$ that renders a uphill hydrogen diffusion and an expedited electronic phase transition through hydrogenation.[31] In addition, the formation energy and thermodynamical migration barrier for oxygen vacancies in perovskite oxide system are pronouncedly reduced under *in-plane* tensile distortion, in favor of a higher concentration of oxygen defects.[24, 25, 32-34] Therefore, the acceleration in defect-engineered electronic phase transitions of correlated VO$_2$ system through strain control may be attributed to the reduction in the formation and diffusion barrier for oxygen vacancies via *in-plane* tensile distortion. This overwhelming advantage associated with interfacial strain offers a fertile ground to flexibly adjust defect-engineered phase transitions in correlated VO$_2$ system, which is poised to overcome the primary bottleneck for high-speed iontronic devices.

The variations in the chemical environment of the as-grown $VO_2$ films through oxygen deficiency was investigated by using X-ray photoelectron spectroscopy (XPS), with the results presented in Figures 3d-3e. All the obtained XPS data were first calibrated against the adventitious C 1$s$ peak located at 284.8 eV, while the V 2$p$ peaks were subsequently deconvoluted into the $V^{3+}$ and $V^{4+}$ components to analyze the changes in vanadium valence states. It is found that the valence state of vanadium is gradually reduced from +4 towards +3 via oxygen deficiency, an effect observed in both strain-relaxed and tensile-strained $VO_2$ films, consistent with the electron donation arising from oxygen vacancies. Furthermore, the XPS spectrum of oxygen-deficient $VO_{2-x}$ shows a marked enhancement of the oxygen defect peak relative to the V-O lattice oxygen peak, providing spectroscopic signature of effective vacancy incorporation (Figure S7). The elevation in the oxygen defect to lattice oxygen ratio through defect engineering is more pronounced for tensile-strained $VO_2$ than in its strain-relaxed counterpart, a trend directly aligned with the observed structural and electronic phase transitions, further validating the strain-enhanced defect evolution.

To clarify the physical origin behind defect-mediated electronic phase transitions in correlated $VO_2$ system, soft X-ray absorption spectroscopy (sXAS) technique was performed, as their V-$L$ edge and O-$K$ edge shown in Figures 4a-4b, respectively. In Figure 4a, the V-$L$ edge spectrum represents the V 2$p$ → 3$d$ transition in $VO_2$, a sensitive indicator of the variations in the vanadium valence state.[35, 36] Oxygen deficiency in $VO_2$ results in the leftward shift of both V-$L_{III}$ and V-$L_{II}$ peaks that manifests the reduction in the valence state of vanadium, corroborating the XPS result. Given the strong V-3$d$ and O-2$p$ orbital hybridization and empty O-2$p$ states, the O $K$-edge spectrum associated with the O 1$s$ → 2$p$ transition reflects the unoccupied density of states in the conduction band for $VO_2$. Therefore, the first (second) peak intensity can qualitatively represent the electron occupation in the low-energy (high-energy) $t_{2g}$ ($e_g$) band of $VO_2$.[37] The electron carriers introduced by defect engineering preferentially occupy the low-energy $t_{2g}$ band in $VO_2$, as evidenced by the lowering of the $t_{2g}/e_g$ peak intensity ratio. Band filling in the $t_{2g}$ orbital resulting from the defect-associated electron doping reconfigures the electronic band structure of correlated $VO_2$ that drives the Mott phase modulation from electron-localized $t_{2g}^1 e_g^0$ state to electron-itinerant $t_{2g}^{1+\Delta} e_g^0$ state.

Establishing the relationship between microscopic oxygen defect concentration and macroscopic electronic phase modulation benefits the understanding of how oxygen vacancies behave in defect-induced phase transition. It is worth noticing that hydrogen-associated electron-doping Mottronic phase modulations in $VO_2$ system are intimately related to the incorporated hydrogen concentration and resultant band filling, engendering opposite orbital reconfigurations towards either electron-localized state or electron-itinerant state.[38] Prolonging the high-vacuum annealing duration from 10 min to 1 h results in a progressive lattice expansion and sequential electronic phase transition in $VO_2$ through oxygen defects (Figures 4c-4d). Elevating the oxygen defect concentration in $VO_2$ through extending the annealing period enhances the electron filling in the $d_{//}^*$ orbital that triggers the more complete carrier delocalization, unveiling band-filling-mediated phase transformation. Consequently, the physical origin associated with defect-mediated electronic phase transitions in $VO_2$ system is identified as electron doping in $t_{2g}$ orbital, triggering Mottronic orbital configuration from electron-localized $t_{2g}^1 e_g^0$ state to electron-itinerant $t_{2g}^{1+\Delta} e_g^0$ state that can be further accelerated by imparting interfacial strains (Figure 4e). Traditionally, thermally-triggered IMT property of $VO_2$ is driven by the splitting of $d_{//}$ orbital and/or the V-V dimerization, simultaneously accompanied by the structural transformation (Figure S8). In particular, the competing instabilities in multiple degrees of freedom results in an exceptionally sensitive electronic phase diagram of correlated $VO_2$ that can facilely tuned by using bandwidth and band-filling control. Our findings unveil the unique advantage of interfacial strain in controlling the IMT property of $VO_2$ through two distinct mechanisms: 1) precise bandwidth control for tuning thermally activated IMT; 2) lowering energy barriers for oxygen defect dynamics to facilitate filling-controlled phase modulations.

## 4. Conclusion

In this work, we showcase the precise control over thermally-driven and defect-mediated electronic phase transitions in correlated $VO_2$ system through interfacial strains. The *in-plane* tensile distortion imposed by lattice mismatching between $VO_2$ film and $TiO_2$ substrate triggers an enhanced orbital hybridization and a wider bandwidth, resulting in a tunable IMT functionality across a temperature range of 291-341 K via bandwidth control. The oxygen deficiency introduces electron carriers into the $t_{2g}$ band of $VO_2$ that triggers Mottronic orbital configuration from electron-localized $t_{2g}^1 e_g^0$ state to electron-itinerant $t_{2g}^{1+\Delta} e_g^0$ state that correlates with the incorporated defect concentration. Such the band-filling control over structural and electronic state evolution in correlated $VO_2$ can be readily facilitated by imparting *in-plane* tensile strain via modifying respective defect dynamics. The overwhelming advantage of interfacial strain not only establishes a unified framework for IMT control in correlated oxide system by integrating bandwidth and band-filling mechanisms, but also paves the way for advanced high-speed iontronic devices via adjusting defect dynamics.

**Notes**

The authors declare no competing financial interest.

**Acknowledgements**

This work was supported by the National Natural Science Foundation of China (No. 52401240), Fundamental Research Program of Shanxi Province (No. 202403021212123), and Scientific and Technological Innovation Programs of Higher Education Institutions in Shanxi (No. 2024L145). The authors also acknowledge the beam line BL08U1A at the Shanghai Synchrotron Radiation Facility (SSRF) (https://cstr.cn/31124.02.SSRF.BL08U1A) for the assistance of sXAS measurement.

**Additional information**

Supporting Information is available online.

**Author contributions**

X.Z. conceived this study, and lead the project; X.Y. and X.Q. grew $VO_2$ films, and carried out the transport measurements under the supervision of X.Z.; X.Z. analyzed the results and wrote the paper; All the authors discussed the results and commented on the final manuscript.

**Correspondences:** Correspondences should be addressed: Dr. Xuanchi Zhou (*xuanchizhou@sxnu.edu.cn*).

**Figures and captions**

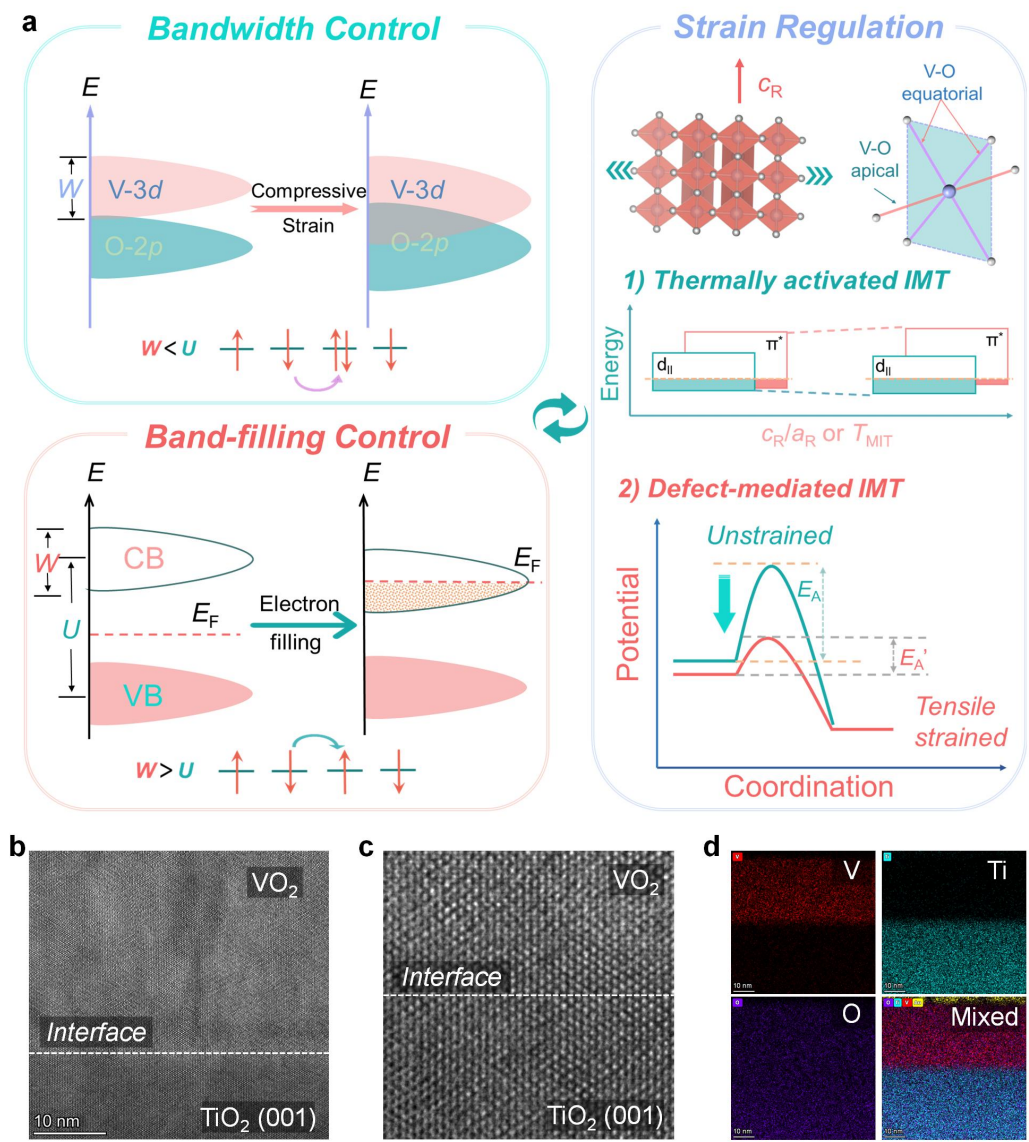

**Figure 1. Manipulating the insulator-metal transitions of VO$_2$ using interfacial strains. a,** Schematic diagram of adjusting the electronic phase transitions in correlated VO$_2$ through bandwidth and band-filling control. **b,** The low-magnification cross-sectional high-resolution transmission electron microscopy (HRTEM) images for the VO$_2$ films deposited on *c*-plane TiO$_2$ substrate. **c,** Cross-sectional HRTEM images for the interfacial region of the grown VO$_2$ films on the *c*-plane TiO$_2$ substrate. **d,** Energy dispersion spectrum (EDS) elemental mapping of vanadium, titanium, and oxygen elements in the as-grown VO$_2$/TiO$_2$ (001) heterostructure.

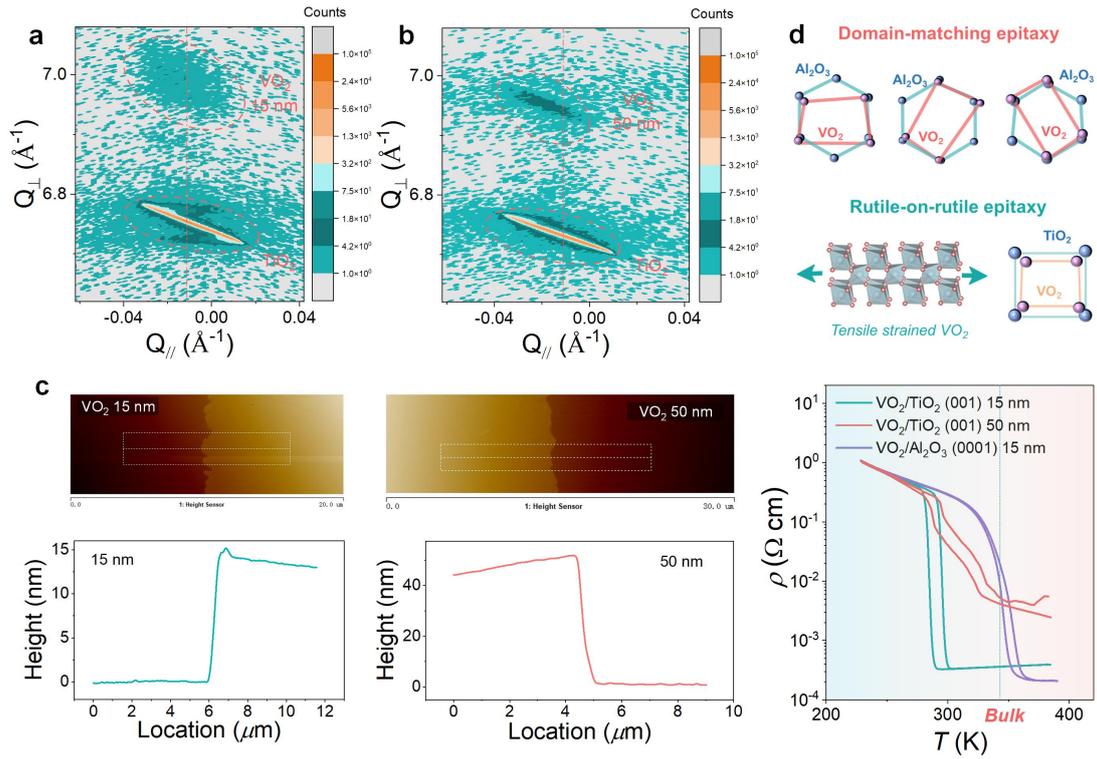

**Figure 2. Artificial design in interfacial strain states of the grown VO$_2$ films.** Reciprocal space mapping (RSM) spectra for **a,** 15 nm and **b,** 50 nm-thick VO$_2$ films deposited on (001)-oriented TiO$_2$ substrates. **c,** Thickness mapping of the VO$_2$/TiO$_2$ heterostructure determined by using atomic force microscope (AFM). **d,** Temperature dependence of material resistivity ($\rho$-$T$) as compared for VO$_2$/TiO$_2$ (001) and VO$_2$/Al$_2$O$_3$ (0001) heterostructures.

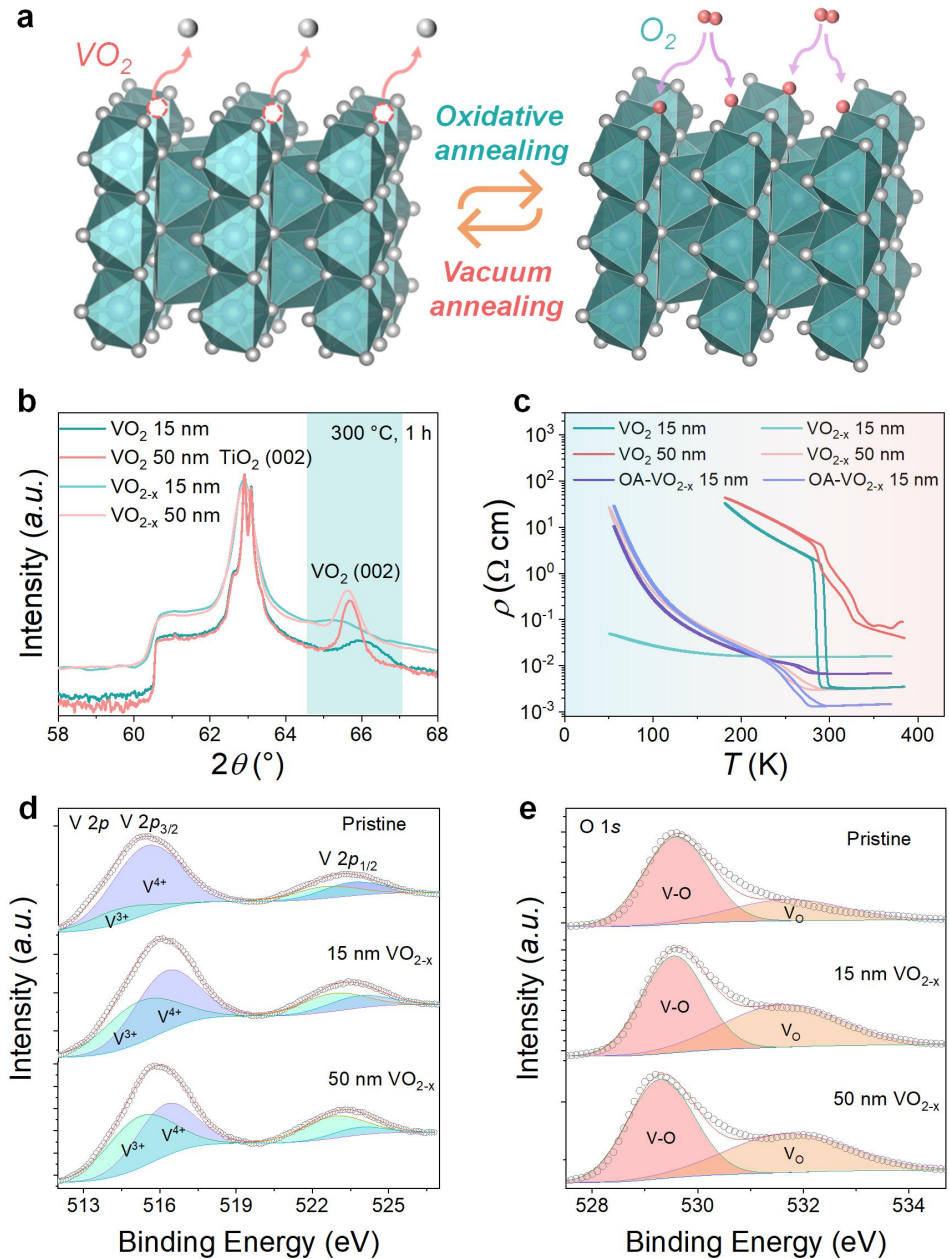

**Figure 3. Tailoring defect-engineered electronic phase transitions of VO$_2$ via interfacial strains. a,** Schematic illustration of the reversible oxygen adsorption and desorption process in VO$_2$ thin film. **b,** X-ray diffraction (XRD) patterns compared for VO$_2$ with different interfacial strain states through oxygen defects. **c,** $\rho$-$T$ tendencies as measured for oxygen-deficient VO$_{2-x}$ with different status of interfacial strains, while oxidative annealing is denoted as OA. **d-e.** X-ray photoelectron spectra (XPS) compared for the core levels of **d,** vanadium and **e,** oxygen of oxygen-deficient VO$_{2-x}$ under interfacial strains.

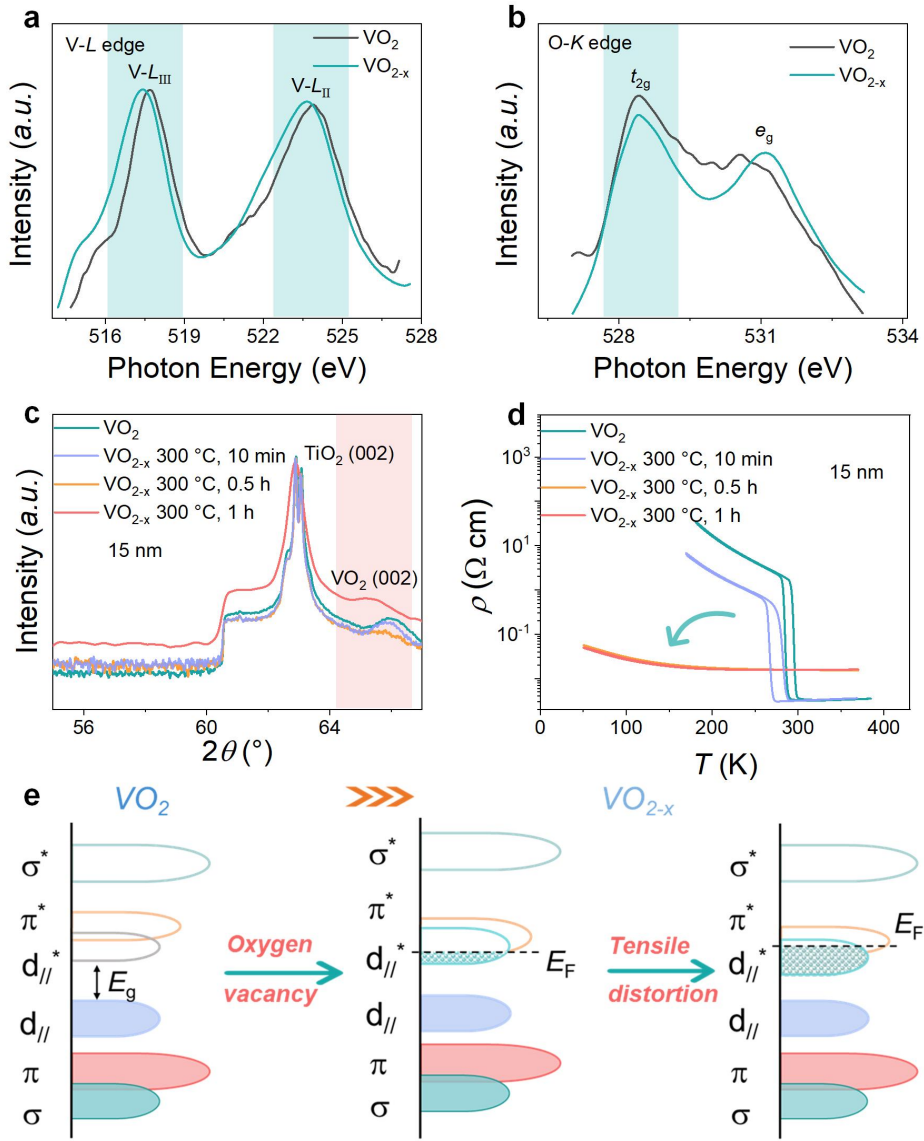

**Figure 4. Electronic band structure for VO₂ through defect engineering.** Soft X-ray absorption spectra (sXAS) for the **a**, V-*L* edge and **b**, O-*K* edge of oxygen-deficient VO$_{2-x}$. **c-d,** Comparing the **c,** XRD patterns and **d,** $\rho$-$T$ curves for oxygen-deficient VO$_{2-x}$ through different annealing durations. **e,** Schematic diagram illustrating the electronic orbital reconfiguration of oxygen-deficient VO₂ via imparting interfacial strains.